\documentstyle[aps,psfig,multicol,eqsecnum,epsf,amssymb]{revtex}
\begin{document}
\title{Algorithm for normal random numbers}
\author{Julio F. Fern\'andez$^{1}$ and Carlos Criado$^2$}
\address{$^{(1)}$Instituto de Ciencia de Materiales de Arag\'on\\Consejo
Superior de Investigaciones Cient\'{\i}ficas\\and Universidad de Zaragoza,
50009-Zaragoza, Spain\\$^{(2)}$Departamento de F\'{\i}sica Aplicada I,
Universidad de M\'alaga\\ 29071-M\'alaga, Spain\\}   
\date{\today}
\maketitle
\vspace{0.3cm} 
\begin{abstract}                
We propose a simple algorithm for generating normally distributed pseudo random
numbers. The algorithm simulates $N$ molecules that exchange energy
among themselves following a simple stochastic rule.
We prove that the system is ergodic, and that
a Maxwell like distribution that may be used as a source of
normally distributed random deviates follows in the $N\rightarrow\infty$ limit.
The algorithm passes various performance tests, including Monte
Carlo simulation of a finite 2D Ising model using Wolff's algorithm.
It only requires four simple lines of computer code, and
is approximately ten times faster than the Box-Muller algorithm.
\end{abstract}  
\pacs{02.70.Rw}
\twocolumn
\tightenlines

Pseudo random number (PRN) generation is a subject of considerable
current interest\cite{reviews}. Deterministic algorithms lead
to undesirable correlations, and some of them have been shown to give rise
to erroneous results for random walk simulations \cite{rw},
Monte Carlo (MC) calculations \cite{ising1,clock}, and growth models \cite{growth}.
Most of the interest has been focused on PRN's with uniform distributions.
Less attention has been paid to non-uniform PRN generation.

Sequences of random numbers with Gaussian
probability distribution functions (pdf's) are needed to simulate
on computers gaussian noise that is inherent to a wide variety of
natural phenomena \cite{noise}. Their usefulness transcends physics.
For instance, numerical simulations of economic systems that make use
of so called {\it geometric} Brownian models (in which noise is multiplicative)
also need a source of normally distributed PRN's \cite{econo}.
There are several algorithms available for PRN's with Gaussian pdf's\cite{devroye}.
Some, such as Box-Muller's algorithm, require an input of uniform PRN's,
and their output often suffers from the pitfalls of the latter \cite{BM}.
Robustness is therefore a relevant issue.
In addition, Box-Muller's algorithm is slow
and can consequently consume significant
fractions of computer simulation times \cite{jffjr}. The comparison
method demands several uniform PRN's per normal PRN, and is therefore
also slow \cite{comparison}. Use of tables \cite{toral} is not a
very accurate method. Algorithms that are related, but not equivalent,
to the one we propose here have been published \cite{jffjr,wallace},
but they are somewhat cumbersome to use. In addition, no proof of their validity
has been given.

We propose here a new algorithm for the generation of normally
distributed PRN's that is quite simple and fast. It
is a stochastic caricature of a closed classical system of $N$ particles. Their
velocities provide a source of PRN's. We prove that, for any initial state,
their pdf becomes Maxwellian in the
$N\rightarrow\infty$ limit, after an infinite number of 
two-particle ``collisions'' take place. To this end, we first prove
that our system is ergodic \cite{lebow,arnold}. The proof is not exceedingly
difficult because our system is not deterministic. We also study
its output as a function of $N$, and establish
useful criteria for its implementation. Correlation test results
are also reported.

For the motivation,
consider numbers $v_1,v_2 \ldots v_N$, placed in $N$ computer registers,
analogous to velocities of $N$ particles that make up
a closed classical system in 1D. Pairs of registers $\imath$ and $\jmath$, say,
selected at random without bias, are to ``interact'' somehow, conserving
quantity  $v_i^2+v_j^2$. By analogy with the
approach to equilibrium (i.e., to Maxwell's velocities distribution)
that is believed to take place in Statistical Physics, we expect that
sufficient number of iterations will lead to an approximately
Gaussian pdf of register values, from which the desired PRN's may be drawn.
(See also Ref. \cite{jffjr}.) 
We define below the simplest interaction we can think of in order that (1)
implementation on a computer be very fast, and (2) that
we may be able to prove that a Gaussian pdf does indeed ensue.

\begin{center}
\begin{figure}
\psfig{file=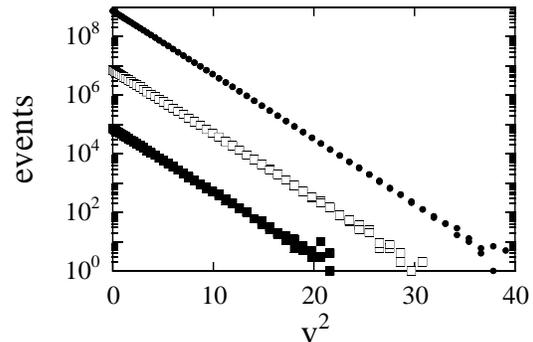,width=8cm}
\caption{Number $n(v)$ of PRN's generated within
$v-\Delta v/2$ and $v+\Delta v/2$, for $\Delta v=0.1$. Datapoins
shown as $\bullet$, $\Box$,
and $\blacksquare$, follow from sequences of $1.9\times 10^{10}$,
$1.8\times 10^8$, and $2\times 10^6$ PRN's, respectively.}
\label{Fig1}
\end{figure}
\end{center}

Before the algorithm is implemented, all $N$ registers must be initialized to,
say, $v_\imath=1$ for all $\imath$ satisfying 
$1\leq \imath\leq N$, or
all $v_\imath$ may be read from a set of $N$ register
values saved from a previous computer run, which we assume to fulfill
$\sum v_\imath^2=N$. Let $U(1,N)$, $U_\imath(1,N)$ be unbiased integer random variables,
both in the interval $[1,N]$, except that $U_\imath$ cannot equal $\imath$.
The algorithm follows:
\begin{eqnarray}
\imath=U(1,N);\;\jmath=U_i(1,N);\\
\label{1}
v_\imath\leftarrow (v_\imath+v_\jmath)/\sqrt{2};\\
\label{2}
v_\jmath\leftarrow -v_\imath+\sqrt{2}v_\jmath
\label{3}
\end{eqnarray}
The updated value of $v_\imath$, from Eq. (2), is used in Eq. (3).
After an initial {\it warm up} phase (see below), $v_\imath$ and $v_\jmath$
may be drawn each time transformation (1-3) is applied.
They are two independent PRN's, each one with an approximately Gaussian pdf,
with $\langle v_\imath\rangle =0$ and $\langle v^2_\imath\rangle =1$
for all $\imath$, if $N$ is sufficiently large (see below).
Transformation (1-3) may be thought of as a rotation of $\pm\pi/4$
with respect to a randomly chosen $\imath \jmath$
plane ($+$ and $-$ signs are for the two possible index orderings,
$\imath \jmath$ and $\jmath \imath$). Thus, quantity $\sum v_{\ell}^2$ is conserved.
Frequencies of events from sequences of $10^6$, $10^8$ and $10^{10}$
PRN's generated with transformation (1-3), with $N=1024$, are exhibited
in Fig.\ref{Fig1}.

We first explain why PRN's genered by transformation (1-3) are
expected to be normally distributed.
Let ${\bf P}_n({\bf v})$ be the probability density
at ${\bf v}=(v_1, v_2,\ldots,v_N)$, after transformation (1-3) has
been applied $n$ times, on the $(N-1)$-dimensional
spherical surface ${\cal S}_{N-1}$, of radius $\sqrt N$ given by 
$N=\sum_{\ell=1,\ldots,N} v_\ell^2$.
Let the single register pdf
$p(v)$ be the $n\rightarrow\infty$ limit of $p_n(v)$, where
$p_n(v_1)=\int {\bf P}_n({\bf v})\:dv_2dv_3\ldots dv_N$.
We show further below that ${\bf P}_n({\bf v})\rightarrow constant$ over 
${\cal S}_{N-1}$ as $n\rightarrow\infty$. It then follows by integration that,
\begin{equation}
p(v)\propto \left(1-{{v^2}\over N}\right)^{(N-3)/2}.
\label{max}
\end{equation}
Clearly, $p(v)\rightarrow C\exp(-v^2/2)$ in the $N\rightarrow\infty$ limit,
which is the desired result.

We prove below, in three stages, that $P_n({\bf v})$ does indeed become
homogeneous over spherical surface ${\cal S}_{N-1}$, if $N\geq 3$, in the
$n\rightarrow\infty$ limit. We first prove $P_n({\bf v})\leftrightarrow
P_n({\bf u})$ as $n\rightarrow\infty$ if ${\bf v}$ and ${\bf u}$ are related.
[From here on, we say that points ${\bf v}$ and ${\bf u}$ are {\it related}
if succesive transformations (1-3) of ${\bf v}$ can lead to ${\bf u}$.]
We then prove that the system's ``orbit'' covers ${\cal S}_{N-1}$ densely 
[that is, that any point ${\bf v}\in{\cal S}_{N-1}$ can be brought arbitrarily
close to any other point ${\bf u}\in{\cal S}_{N-1}$ by applying
transformations (1-3) to ${\bf v}$ a sufficient number of times].
Then, the desired result follows easily. It may help to place the
significance of the proof that follows into proper perspective to note that
if in Eq. (1) $\jmath\leftarrow U_i[1,N]$ is replaced by
$\jmath=\imath +1\; mod N$,
the system becomes then non-ergodic, as can be easily checked numerically.

To start the proof, let kernel $K({\bf v},{\bf v^\prime})$ be defined
by
$P_{n+1}({\bf v})=\int K({\bf v},{\bf v^\prime})P_n({\bf v^\prime})\;d{\bf v^\prime}$,
and let 
\begin{equation}
F_n\equiv \int \{P^2_{n+1}({\bf v})-P^2_n({\bf v})\}\;d{\bf v}.
\end{equation}
Note first that $F_n<0$ implies that $P_{n+1}({\bf v})$ is more uniform than
$P_n({\bf v})$, in the sense that
$\int d{\bf v}\;[P_{n+1}({\bf v})-{\overline P}]^2
<\int d{\bf v}\;[P_{n}({\bf v})-{\overline P}]^2$,
where ${\overline P}=1/\int dv $.
It follows from the definition of $K({\bf v},{\bf v^\prime})$ that
\begin{equation}
F_n=
\int d{\bf v}\{[\int d{\bf v_1}K({\bf v},{\bf v}_1)
P_n({\bf v}_1)]^2-P^2_n({\bf v})\}.
\label{Fn}
\end{equation}
Making use of the detailed balance condition,
$K({\bf v},{\bf v^\prime})=K({\bf v^\prime},{\bf v})$, which our system
satisfies, and the relation $\int\;d{\bf v}\;K({\bf v},{\bf v^\prime })=1$,
Eq. (\ref{Fn}) can be cast into,
\begin{equation}
F_n=
-{1\over 2}\int d{\bf v}\int d{\bf v_1}\int d{\bf v_2}\;
Q({\bf v},{\bf v}_1,{\bf v}_2),
\end{equation}
where, $Q=K({\bf v},{\bf v}_1)
K({\bf v},{\bf v}_2)[P_n({\bf v}_1)-P_n({\bf v}_2)]^2$.
Therefore, in the $n\rightarrow\infty$ limit, $P_n({\bf v})$
becomes constant over each set in ${\cal S}_{N-1}$
within which any two points ${\bf v,u}$ are
related.

We now prove that the system's orbit covers ${\cal S}_{N-1}$ densely.
Let $H_N$ be the group of transformations
in $N$ dimensions defined by Eqs. (1-3). We first show that any
rotation in $3$D [that is, any element of $SO(3)$]
can be approximated arbitrarily close by elements of $H_3$.
The proof is extended to higher dimensions by induction.
Note first that $H_3$ does not belong to the set of {\it finite}
rotation groups in $3$D\cite{coxeter}, and is therefore
an infinite group. Let group $SO(3)$ be covered by spheres of
radius $\epsilon/2$ each. A finite number of them is sufficient,
since the volume of $SO(3)$ is finite \cite{gilmore}. It follows that
there must be at least one sphere with two elements of $H_3$ in it,
since $H_3$ has an infinite number of elements. Let these two elements
be $r$ and $s$, and let $g({\bf u},\epsilon )$ be element $r s^{-1}$
of $H_3$, which is a rotation by angle $\epsilon$ about some
undetermined ${\bf u}$ axis. We will build elements of $H_3$ that
are as near as desired to any given rotation. To this end, it is sufficient to show
that it can be done for a set of infinitesimal generators of rotations
\cite{goldstein}. One such set is made up of infinitesimal rotations about
three linearly independent axes. Consider axes ${\bf u}_1$, ${\bf u}_2$,
and ${\bf u}_3$ that are obtained from ${\bf u}$ by rotations $g(1,\pi /2)$,
$g(2,\pi /2)$, and $g(3,\pi /2)$ about each one of the coordinate axes by
angle $\pi /2$. The correspondng infinitesimal rotations are given by
\cite{gilmore},
$g({\bf u}_\imath,\epsilon )=g(\imath ,\pi /2) g({\bf u},\epsilon)
g^{-1}(\imath ,\pi /2).$
This concludes the proof for 3 dimensions. 

We now prove by induction that any element $g(\imath \jmath ,\alpha )$,
for the rotation about plane $\imath \jmath$, by angle $\alpha$ of 
the rotation group $SO(N)$ can be approximated as nearly as desired
by an element $g$ of group $H_N$, for $N>3$. By hypothesis, any
$g(\imath \jmath ,\alpha )$, for $\imath ,\jmath =1,2,\ldots ,N$ can be
approximated by an element $g$ of $H_N$. We show now that
$g(\imath\; N+1 ,\alpha )$, for $\imath =1,2,\ldots ,N$,
can also be approximated by elements of $H_{N+1}$. We take $g\in H_N$
within distance $\epsilon$ of $g_{\imath \jmath}(\alpha )$. Now, since
rotations preserve distances, it follows that $g(\imath\;N+1,\alpha )\in SO(N+1)$,
given by
$g(\imath\;N+1,\alpha )=g(\imath\;N+1,\pi /2)
g(\imath \jmath ,\alpha )g^{-1}(\imath\;N+1,\pi /2)$
is within distance $\epsilon$ of $g^\prime\in H_{N+1}$, given by
$g^\prime =g(\imath\;N+1,\pi /2)
gg^{-1}(\imath\;N+1,\pi /2)$. This proves dense coverage in $N\geq 3$ dimensions.
This is a stochastic generalization of Jacobi's theorem \cite{arnold}
to more than two dimensions.

To conclude the proof that $P_n({\bf v})\rightarrow constant$ in the
$n\rightarrow \infty$ limit, consider any two points
${\bf V}$ and ${\bf U}^\prime$ as centers
of disks ${\cal D}_{\bf V}$ and ${\cal D}_{{\bf U}^\prime}$, both
of radius $r$, in ${\cal S}_{N-1}$. Since the system's orbit covers
${\cal S}_{N-1}$ densely for $N\geq 3$, it follows that
a point ${\bf U}$ that is related to ${\bf V}$
exists arbitrarily close to ${\bf U}^\prime$. Consider now disks
${\cal D}_{\bf V}$ and ${\cal D}_{\bf U}$. The fact that there exists at least
one sequence of rotations in $H_N$ that take ${\bf V}$ into ${\bf U}$ implies that
there exists at least one single rotation $g$ in $H_N$ that transforms ${\bf V}$
into ${\bf U}$. Since $g$ is a rotation, it transforms ${\cal D}_{\bf V}$
{\it rigidly} into ${\cal D}_{\bf U}$. It follows that $\int d{\bf v}\; P_N({\bf v})$
over ${\cal D}_{\bf V}$ equals $\int d{\bf u}\;P_N({\bf u})$ over
${\cal D}_{\bf U}$. Since $r$ is arbitrary, and ${\bf V}$ and ${\bf U}^\prime$
are any two points in ${\cal S}_{N-1}$,
it follows that $P({\bf v})$ is constant over ${\cal S}_{N-1}$ (except, perhaps,
on a set of measure zero). This is the desired result. Ergodicity follows
\cite{arnold}.  

We next address the following practical issues: (1) how good an approximation
to a Gaussian pdf of PRN's is achieved with a necessarily
{\it finite} set of $N$ registers; (2) how long must the warm up phase be.
 
It is convenient to rewrite Eq. (\ref{max}) as follows,   
\begin{equation}
p(v)\propto e^{-v^2/2}e^{{g_N(v)}/{N}}.
\label{gofx}
\end{equation}
where $g_N(v)=v^2(3-v^2/2)/2+{\cal O}(1/N)$. 
$N^{-1}g_N(v)$ is approximately the fractional deviation,
$\delta p(v)/p(v)$, from Gaussian
form if $\delta p(v)/p(v)\ll 1$. We have checked this
behavior numerically. 
Clearly, the {\it number of registers} $N$ that must be used increases with
the number $M$ of PRN's one intends to generate. This is because
the value of the largest PRN generated increases, on the average, with $M$.
More precisely, the value of $v$ beyond which PRN's are only generated with
probability $q$ is approximately given by $v^2\approx 2\ln(M/vq)$.
Now, it follows from Eq. (\ref{gofx}) that the fractional error $\delta P/P$
in the probability density at $v$ is approximately $N^{-1}v^2(3-v^2/2)/2$
for very large $N$. (It is pointless to require this error to be too
small since a PRN is expected to be generated beyond $x$
with a small probability $q$.)
It then follows that $[\ln(M/qv)]^2\lesssim N\delta P/P$ must be satisfied by $N$.
Thus, approximately $10^4$ registers are sufficient
in order to generate as many as $10^{15}$ PRN's,
with a roughly $10\%$ error in the probability for the
{\it largest} PRN in the sequence. For results obtained from a sequence
of $10^{10}$ PRN's generated with $1024$ registers, see Fig. 1.
\begin{center}
\begin{figure}
\psfig{file=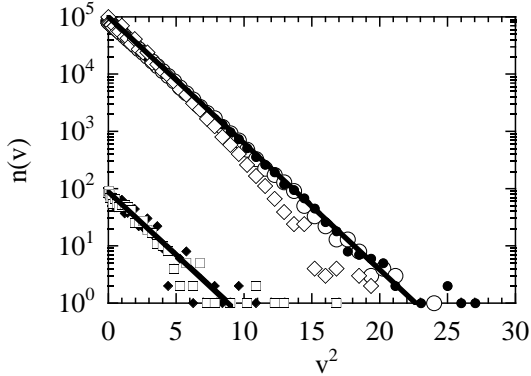,width=8cm}
\caption{Number  $n(v)$ of PRN's within
$v-\Delta v/2$ and $v+\Delta v/2$, for $\Delta v=0.1$,
starting from initial conditions $v_\imath =1$, for all $\imath\in [1,N]$, after
transformation (1-3) is iterated
$2n_pN$ times (that is, after each register interacts, on the average,
$2n_p$ times). The $\Box$ and $\blacklozenge$ stand for $n_p=2,10$,
respectively, for $N=1024$. The $\Diamond$ and $\bullet$,
and $\circ$ stand for $n_p=2,4$, and $10$ respectively, for $N=1 048 576$.
The two straight lines stand for $C\exp (-v^2/2)$ for two values of $C$.}
\label{Fig2}
\end{figure}
\end{center}

Our algorithm must be applied a number $n_pN$ of times before it is
ready for use unless all $v_\imath$ are initialized with``equilibrium''
values (stored from some previous computer run). The distribution
of all register values then evolves towards equilibrium, as illustrated
in Fig. 2. Deviations from equilibrium are statistically
insignificant for $n_p\gtrsim 2$ and $N=1024$,
and for $n_p\gtrsim 4$ and $N=1\:048\:576$. 
Since $n_p$ is expected to increase as $\ln N$, $n_p=8$ should
provide ample warm up for any forseeable applications.

The number of PRN's that must be generated before each PRN in
sequence $v_1,v_2,\ldots,v_N$
returns within distance $r$ from its initial value is exponential in $N$. More
specifically, we estimate it to be
$(\tau /\sqrt{N})(1/r)^N$ for $N\gg 1$, where
$\tau$ is the period of the algorithm used to select $\imath$
and $\jmath$ in Eq. (1). The estimation is based on
$P_n({\bf v})\rightarrow constant$ over ${\cal S}_{N-1}$ as $n\rightarrow\infty$.
Thus, an effectively infinite recurrence time follows for any reasonable value of $N$. 

Correlations between a finite number of PRN's
clearly vanish as $N\rightarrow\infty$, since $\imath$
and $\jmath$ in Eq. (1) are supposedly independent PRN's. We have searched for
correlations in $m$ succesively generated PRN's $v_1,v_2,\ldots v_m$,
for $m=3,4,\ldots ,6$, performing a chi-square isotropy test over
the corresponding $m$-dimensional space. An $m$-tuple
${\bf v}=v_1,v_2,\ldots,v_m$ was said to belong to
the $i-th$ cone, of $1024$ randomly oriented
cones with axes ${\bf w}_1,{\bf w}_2,\ldots,{\bf w}_{1024}$, if 
$0.99\leq {\bf v.w_\imath}\leq 1$.
No significant deviations from isotropy were observed for 
$10^6$ generated $m$-tuples.
\begin{center}
\begin{figure}
\psfig{file=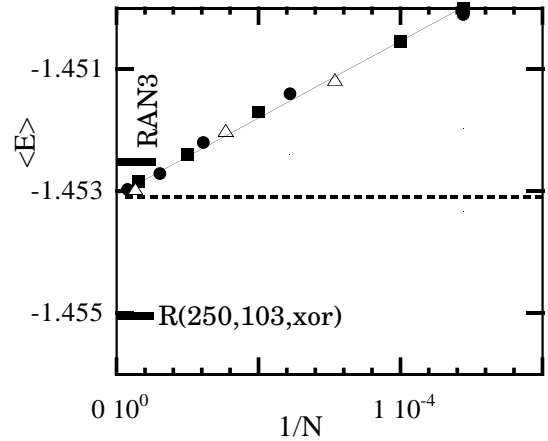,width=8.5cm}
\caption{Average energy per spin, obtained from MC
simulations using Wolff's algorithm, versus the inverse of the number of registers
used for the generation of PRN's with Gaussian pdf's. $\bullet$, $\blacksquare$,
and $\vartriangle$ stand for data points that follow from feeding our algorithm
with the following uniform PRN generators: ggl, R(250,103,xor), and
Ran3, respectively. Unacceptable energy values that have been obtained
in Refs. [3] using R(250,103,xor), and Ran3 
are also shown as bars next to the $y$-axis.}
\label{Fig3}
\end{figure}
\end{center}
 
Implementation of Wolff's algorithm \cite{wolff} in MC
calculations of the Ising model's critical behavior is a demanding
test that some well known uniform PRN generators have failed \cite{ising1}.
Large clusters are then flipped as a whole, and this
tests correlations in very long sequences. We have used
normal PRN's generated by our algorithm as input into a MC
simulation of an Ising system of $16\times 16$ 
spins at the critical temperature. [For that, we note that
$v^2_\imath +v^2_\jmath >2x$ as often as $u>\exp (-x)$ if $v_\imath$ and
$v_\jmath$ ($u$) are PRN's with Gaussian (uniform) pdf's, respectively.]
The energy obtained is shown in Fig. 3 as a function of the number of
registers $N$. The following
uniform PRN algorithms were used to select $\imath$ and $\jmath$ in Eq. (1):
ggl \cite{ising1}, R(250, 103,xor)
\cite{rw,ising1}, and RAN3 \cite{nr}. We tried the latter two algorithms,
which have been shown to lead by themselves to unacceptable results for the
Ising model \cite{ising1}, in order to test our algorithm's
robustness. The results shown in Fig. 3 are gratifying.

Similarly, the specific heat $c$ and magnetization $m$ fluctuations
data points obtained follow approximately the relations $c\simeq c_0+8.4/N$, and
$\langle (\delta m)^2\rangle \simeq \chi_0+33/N$, respectively,
where $c_0=1.497(1)$ and $\chi_0=0.5454(2)$, in agreement with the
known exact values\cite{ising1,fisher}. 

Double precision is recommended. It prevents excesive drift of the sum
$\sum v_\imath^2$ away from its assigned value.
Even then, single precision accuracy is to be expected at the end
of a sequence of some $10^{16}$ PRN's, unless the sum is normalized
several times during the run.

In summary, we have shown that implementation of Eqs. (1-3) provides
a source of PRN's with an approximately Gaussian pdf.
Some $10^4$ registers (molecules)
are sufficient for some purposes, but up to $10^5$ or more may be necessary
for more demanding tasks. (Having to make a decision about the number of registers to
be used may sometimes be an unwelcomed task. On the other hand,
it is a virtue of the algorithm, that one can control, through the value of $N$,
how close the output is to be from sequences of truly independent random numbers
with Gaussian pdf's.) Initial warm ups for arbitrary initial conditions are
necessary; it is sufficient to let each register initially interact an
average number of, say, 8 times. The system's recurrence time was shown to be
exponential in $N$, and therefore effectively infinite. Its behavior
appears to be robust. The proposed algorithm runs an order of magnitude faster
on computers than the most often used Box-Muller method \cite{devroye,BM}. 
For a fortran code of our algorithm or other questions,
please write JFF@Pipe.Unizar.Es.

Continuous help from Dr. Pedro Mart\'{\i}nez with computer systems
is deeply appreciated by JFF. We are indebted to 
Prof. P. Grassberger for an important suggestion.
JFF and CC are grateful for partial financial support from
DGICYT of Spain, through grants No. PB95-0797 and PB97-1080, respectively.

\end{document}